\documentstyle[11pt,newpasp,twoside,epsf]{article}
\def\edcomment#1{\iffalse\marginpar{\raggedright\sl#1\/}\else\relax\fi}
\reversemarginpar

\begin{document}
\title{Nucleosynthesis of Light and Heavy Elements in Baryon-Rich Outflows Associated with Gamma-Ray Bursts}
\author{Susumu Inoue}
\affil{Max-Planck-Institut f\"ur Astrophysik,
       Karl-Schwarzschild-Str.1, Postfach 1317, 85741 Garching, Germany}
\author{Nobuyuki Iwamoto, Manabu Orito, Mariko Terasawa}
\affil{National Astronomical Observatory, 2-21-1 Osawa, Mitaka, Tokyo 188-8588, Japan}

\begin{abstract}
Gamma-ray bursts (GRBs) must originate from low baryon load, ultrarelativistic outflows;
 however, slower, more baryon-rich outflows (BROs) should also arise
 in connection with GRBs as ``circum-jet winds'' and/or ``failed GRBs''.
We study the possibility of nucleosynthesis within such BROs
 by conducting detailed reaction network calculations
 in the framework of the fireball model,
 showing that they can be interesting sites
 for synthesis of heavy neutron capture elements,
 as well as of light elements such as deuterium.
These products may be observable
 in the companion stars of black hole binary systems or in extremely metal-poor stars,
 offering an interesting probe of conditions in the central engine.
\end{abstract}

\section{Introduction}
Successful generation of GRBs requires the formation of
 an ultrarelativistic outflow with bulk Lorentz factor $\Gamma \ga$ 100,
 implying very low baryon-loading
 (e.g. M\'esz\'aros 2002).
Since the temperature at the base of the outflow should be of order MeV
 and the baryons are likely to contain a high fraction of neutrons
 (Pruet, Woosley \& Hoffman 2003, Beloborodov 2003),
 some production of nuclei starting from free protons and neutrons
 is expected to occur in the expanding flow.
However, as shown in several recent papers (Lemoine 2002, Pruet, Guiles \& Fuller 2002, Beloborodov 2003),
 in the very high entropy
 and extremely rapid expansion
 characteristic of GRB outflows,
 nucleosynthesis is limited to only small production of D and $^4$He,
 which are difficult to observe.

Nevertheless, most GRB progenitors should also give rise to associated outflows
 with higher baryon-loads and lower velocities (baryon-rich outflows, or BROs).
It is highly probable that different types of ``circum-jet winds'' of baryon-loaded material
 surround and coexist with the narrowly collimated GRB jet.
The GRB jet production mechanism is likely to act
 not only on the baryon-poor zones near the jet axis,
 but also on the more baryon-contaminated, peripheral regions,
 leading to BROs as a natural byproduct.
Models involving core collapse and jet penetration in massive stars
 (MacFadyen and Woosley 1999, Wheeler et al. 2002)
 can also induce BROs through entrainment and mixing with the stellar material.
In black hole accretion disk models,
 baryon-rich winds may arise from the outer disk
 by mechanisms distinct from the GRB jet
 (Janka \& Ruffert 2001, Narayan, Piran \& Kumar 2001, MacFadyen 2003).
Some observational evidence support the existence of such circum-jet BROs
 (e.g. Woosley, Zhang \& Heger 2002).
Alternatively, BROs may occur without concomitant GRBs as ``failed GRBs'',
 potentially with event rates higher than successful GRBs,
 if the baryon-loading process acts more thoroughly
 or the the GRB jet-driving mechanism operates less efficiently in the central engine.
These may possibly be identified with observed ``hypernovae'' (Nomoto et al. 2002),
 X-ray flashes (Kippen et al. 2002),
 or with some hitherto unrecognized type of transient.
Either way, the lower entropy, slower expansion and higher ejecta mass of BROs compared to GRB jets
 make them much more interesting from a nucleosynthesis viewpoint.
As a first study of nucleosynthesis in BROs associated with GRBs, we investigate this problem
 utilizing the simplified dynamical framework of the basic fireball model,
 but incorporating detailed nuclear reaction networks including both light and heavy elements.
More details can be found in Inoue et al. (2003).

\section{Nucleosynthesis in GRB-BROs}

We consider a wind fireball, i.e. an adiabatic, thermally-driven
 and freely-expanding outflow (Paczy\'nski 1990).
The key parameters are
 the luminosity $L = 10^{52} L_{52} {\rm erg \ s^{-1}}$,
 the initial radius $r_0 = 10^7 r_{0,7} {\rm cm}$,
 and the dimensionless entropy $\eta = L/\dot M c^2$.
The fireball first undergoes an acceleration phase,
 where the comoving temperature and density decrease exponentially
 on the initial dynamical timescale $t_{d0} \simeq 0.33 {\rm ~msec} \ r_{0,7}$.
After the internal energy is converted to bulk kinetic energy,
 there follows a coasting phase where $T$ and $\rho_b$ decrease as power-laws,
 asymptotically becoming $T = T_0 \eta^{-1} (t/t_{d0})^{-2/3}$
 and $\rho_b = \rho_{b,0} \eta^{-3} (t/t_{d0})^{-2}$.
These formulae are valid for $\eta \ga 1$.
Although $\eta \la 1$ (nonrelativistic expansion) may be realized in some progenitor models,
 here we assume $\eta \ge 2$,
 corresponding to baryon load mass $M_b \le 2.8 \times 10^{-2} {\rm M_\odot} E_{53}$
 for total energy $E= 10^{53} {\rm erg} E_{53}$,
 which may be a reasonable upper limit for relatively baryon-free progenitors
 such as neutron star mergers (Janka \& Ruffert 2001) or supranovae (Vietri \& Stella 1998).
The initial electron fraction $Y_e$
 is taken to be an additional parameter,
 in view of the uncertainties in its expected value (Pruet et al. 2003).
We implement nuclear reaction network codes
 including a large number of light, very neutron-rich nuclei (Terasawa et al. 2001).

Taking fiducial values of $L_{52}=1$ and $r_{0,7}=1$,
 we concentrate on the effect of variations in $\eta$ and $Y_e$.
Although $^4$He is the dominant final species besides remaining free nucleons for all cases,
 interesting amounts of other light elements can also result,
 particularly deuterium, whose final mass fraction can reach $X_D \simeq 0.02$.
 (Fig.1).
For $\eta \ga 20$, the reactions mainly occur during the exponential expansion (acceleration) phase,
 similar to the case of big bang nucleosynthesis.
Contrastingly, for $\eta \la 10$,
 nucleosynthesis proceeds mainly during the power-law expansion (coasting) phase.
When $Y_e \la 0.5$, free neutrons remain abundant in the outflow
 until they start decaying into protons,
 so that D production can proceed by subsequent p(n,$\gamma$)d reactions at very late times.
Other elements such as $^7$Li, $^9$Be and $^{11}$B
 can be generated at levels of $X \sim 10^{-8}-10^{-7}$.

\begin{figure}
\plotfiddle{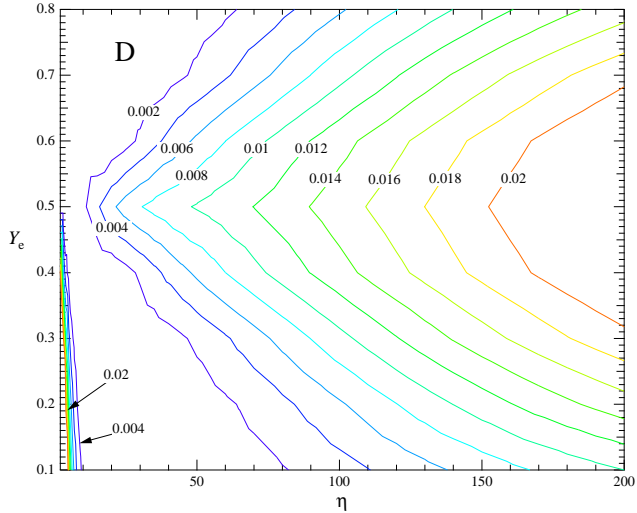}{6cm}{0}{30}{30}{-135}{-15}
\caption{
Contours of final abundance by mass fraction of deuterium
 for different parameter values of $\eta$ and $Y_e$.}
\end{figure}

The synthesis of heavy elements through neutron capture is even more intriguing.
The fiducial $\eta=2$ outflow is characterized by very high entropy ($s/k_{B} \simeq 2500$),
 low density ($\rho_{b,0} \simeq 1.5 \times 10^5 {\rm g~cm^{-3}}$)
 and short initial expansion timescale ($t_{d0} \simeq 0.33$~ms),
 leading to nucleosynthesis which is markedly different
 from the better-studied $r$-process in supernova (SN) neutrino-driven winds (e.g. Qian 2003).
The final abundances are shown in Fig. 2.
When $Y_e \la$ 0.4,
 considerable reaction flows can occur to high mass numbers,
 although the peak abundances are at levels of $Y \simeq 10^{-6}$.
(Note that the apparent peak at $A \simeq$ 250
 is an artifact of our neglect of fission.)
The abundance patterns do not match the solar distribution
 and possess peaks in between the solar $s$- and $r$-process peaks.
This reflects the occurrence of reaction flow paths 
 intermediate between those of the $s$- and $r$-processes,
 resulting from the balance of neutron captures and $\beta$-decays
 rather than (n,$\gamma$)-($\gamma$,n) equilibrium.
Such processes have been discussed in previous contexts and called the ``$n$-process''
 (Blake \& Schramm 1976).
We also note that the neutrons remain abundant after the reactions freeze out
 until they freely decay, instead of being exhausted by neutron captures.

\begin{figure}
\plotfiddle{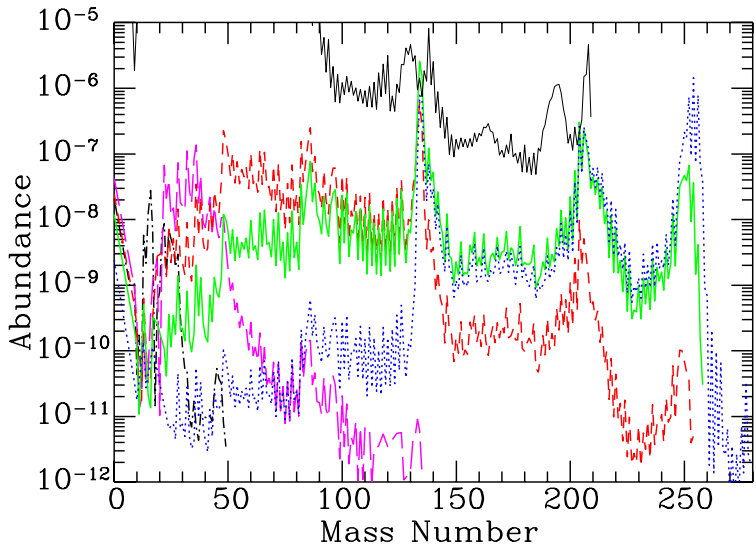}{6cm}{0}{55}{55}{-140}{-10}
\caption{Final abundances as function of mass number,
 for $\eta=2$ and $Y_e =$ 0.1, 0.3, 0.4, 0.48 and 0.498, represented
 by the dotted, solid, dashed, long-dashed and dot-dashed curves, respectively.
The uppermost, thin solid curve is the solar abundance distribution in arbitrary units.}
\label{fig:heavy}
\end{figure}

\section{Observational Implications}
The present results of nucleosynthesis in GRB-BROs have interesting observational implications.
For our case of $\eta=2$ and $Y_e \simeq 0.1-0.4$,
 the ejected mass in heavy elements per single GRB-BRO event should be
 $M_{h,BRO} \sim 10^{-7}$-$10^{-6} {\rm M_\odot}$ for $A \ga 100$,
 comparable to or somewhat less than that expected in $r$-process elements from a normal core collapse SN (Qian 2003).
Considering the fact that the abundance patterns are different from solar,
 and that the intrinsic event rates of (either successful or failed) GRBs 
 are estimated to be less than that for SNe,
 the contribution of GRB-BROs to the solar/Galactic abundances are unlikely to be large,
 at least for the conditions treated here.

However, some observable environments may manifest
 the local abundance pattern of the BRO ejecta.
Many facts suggest that chemical evolution in the early Galaxy proceeded inhomogeneously,
 and also that GRBs are associated with the most massive progenitor stars.
If so, these explosions may have dominated at the earliest epochs, so that
 the BRO nucleosynthetic signature are visible in the most metal-poor stars.
Alternatively, both light and heavy element products
 may be clearly discernible in the companion star of a black hole binary system,
 where the event that formed the black hole may have left traces
 of the explosion ejecta on its surface (Israelian et al. 1999).
High resolution spectroscopic observations of such stars are warranted 
 for detailed studies of their elemental abundance patterns.
Such studies should yield important constraints on the value of $Y_e$,
 providing us with valuable insight into the innermost regions of GRB sources
 which are otherwise very difficult to probe.

\section{Conclusions and Outlook}

In this exploratory study, we have shown that BROs associated with GRBs can be interesting sites
 for nucleosynthesis of both light and heavy elements, with important observational consequences.
We have neglected here various realistic effects that can be potentially important,
 including fission and proton-rich nuclei,
 collimation and higher baryon loading,
 neutrino irradiation, nonthermal spallation, neutron captures on external matter, etc.,
 all of which can be significant, e.g. in the widely discussed collapsar scenario.
Until we arrive at a definitive understanding of the physics of the GRB central engine,
 construction step by step of increasingly realistic models incorporating these effects
 and comparison with various observations
 may help us constrain the unknown properties of the GRB source,
 as well as to clarify the role of GRB-BROs in the chemical evolution of the Galaxy.
It is certain that much interesting work lies ahead.

\end{document}